\documentclass[aps,pre,twocolumn,showpacs,superscriptaddress]{revtex4-1}

\usepackage{amsmath}
\usepackage{amssymb}
\usepackage{graphicx}
\usepackage{lineno}

\begin{document}

\title{Role of functionality in two-component signal transduction: A stochastic study}

\author{Alok Kumar Maity}
\affiliation{Department of Chemistry, 
University of Calcutta, 92 A P C Road, Kolkata 700 009, India}

\author{Arnab Bandyopadhyay}
\affiliation{Department of Chemistry, 
Bose Institute, 93/1 A P C Road, Kolkata 700 009, India}

\author{Pinaki Chaudhury}
\email{pinakc@rediffmail.com}
\affiliation{Department of Chemistry, 
University of Calcutta, 92 A P C Road, Kolkata 700 009, India}

\author{Suman K Banik}
\email{Corresponding author: skbanik@jcbose.ac.in}
\affiliation{Department of Chemistry, 
Bose Institute, 93/1 A P C Road, Kolkata 700 009, India}

\date{\today}

\begin{abstract}
We present a stochastic formalism for signal transduction processes in a bacterial 
two-component system. Using elementary mass action kinetics, the proposed model takes 
care of signal transduction in terms of phosphotransfer mechanism between the cognate 
partners of a two-component system, viz, the sensor kinase and the response regulator. 
Based on the difference in functionality of the sensor kinase, the noisy phosphotransfer 
mechanism has been studied for monofunctional and bifunctional two component 
system using the formalism of linear noise approximation. Steady state analysis of both 
models quantifies different physically realizable quantities, e.g., variance, Fano factor
(variance/mean), mutual information. The resultant data reveals that both systems reliably transfer 
information of extra-cellular environment under low external stimulus and at high kinase 
and phosphatase regime. We extend our analysis further by studying the role of 
two-component system in downstream gene regulation.
\end{abstract}


\maketitle


\section{Introduction}

In response to the changes made in the extra-cellular environment, living systems adapt 
themselves by coordinated regulation of intracellular machinery composed of several
interacting components \cite{Mcadams1995,Tyson2003,Alon2007}. In bacterial kingdom, 
such adaptation is achieved by a group of highly specialized motifs, commonly known 
as two-component system (TCS) \cite{West2001,Laub2007,Kenney2010}. Comprised 
of membrane bound sensor kinase and cytoplasmic response regulator, TCS detects 
changes made in the environment and, in response, controls expression and/or 
repression of one or several downstream genes (target genes other than operon). 
In presence of an external stimulus, 
autophosphorylation takes place in the conserved histidine residue of the sensor 
kinase. The phosphate group is then transferred to its cognate partner, the response 
regulator, containing conserved aspartate domain. When phosphorylated, the response 
regulator regulates one or several downstream genes, as well as its 
own operon. For example, in \textit{M. tuberculosis}, the response 
regulator MprA gets phosphorylated by its cognate sensor MprB in presence of the signal 
and exerts a positive feedback on its own operon \textit{mprAB} \cite{Tiwari2010}. In 
addition to being the source of phosphate group, sensor kinase sometimes 
can dephosphorylate phosphate group from a response regulator by acting as 
phosphatase. This combined kinase and phosphatase activity of the sensor kinase 
makes the TCS bifunctional \cite{Hsing1998,Laub2007,Kenney2010,Goldberg2010}. 
Due to the opposing (kinase and phosphatase) effect of sensor kinase on response 
regulator, bifunctional systems have been placed in a broad category of functional motifs 
known as \textit{paradoxical components} \cite{Hart2012,Hart2013}. In certain TCS, role 
of sensor as a phosphatase is absent, and the job of dephosphorylation is done by an 
auxiliary protein (phosphatase),  thus making the TCS monofunctional 
\cite{Laub2007,Kenney2010}.

Depending on the nature of extra-cellular stimulus, a single bacterium may utilize 
different types of TCS with highly specific functionality to transduce the changes made 
in the surroundings. To sense and adapt appropriately, a single bacterium may contain 
both monofunctional and bifunctional TCS \cite{Laub2007,Kenney2010}. For example, 
\textit{E. coli} chemotaxis system has CheA/CheY TCS that responds to change 
in the chemical gradient in the surrounding where the sensor kinase CheA is monofunctional 
in nature
(acts as kinase only) and role of phosphatase is played by CheZ which is not a part of the 
TCS. On the other hand, EnvZ/OmpR TCS in \textit{E. coli} responds to change in the 
osmolarity of the environment where sensor kinase EnvZ plays a bifunctional role (acts 
as kinase as well as phosphatase). One of the advantages of a bifunctional system over 
a monofunctional one is that it takes care of \textit{input-output robustness} 
\cite{Shinar2007}. Due to its architecture, the output level of the phosphorylated response 
regulator in a bifunctional system depends only on the input stimulus and is independent of 
other system components. On the other hand, such a robustness criterion does not remain valid 
in a monofunctional system. Thus, in the latter case, in addition to the input stimulus, the output 
level depends also on the level of phosphatase which acts on the phosphorylated response 
regulator.

When the aforesaid signal transduction processes are considered within the single cell 
scenario, role of fluctuations, cellular and/or extra-cellular, cannot be ruled out. Whether 
external or internal, such fluctuations not only affect the gene expression mechanism 
within a cell but also control the signal transduction processes involving post-translational 
modifications that are taking place within a noisy environment. With the advancement of 
experimental techniques that employ single cell measurement, it is now possible to quantify 
different physically realizable quantities like variance, Fano factor (variance/mean), etc., 
of different cellular components\cite{Paulsson2004,Rosenfeld2005,Zaslaver2006,Kaern2005,Silvarocha2010,Eldar2010,Balazsi2011,Munsky2012}. 
In this light, it is thus worthwhile to develop a stochastic formalism to study signal 
transduction processes in bacterial TCS keeping in mind the difference in 
functionality of the sensor kinase. Although deterministic modeling of bacterial signal
transduction machinery is known in the literature \cite{Alves2003,Bandyopadhyay2012,Shinar2007,Igoshin2008,Miyashiro2008,Tiwari2010}, 
few attempts have been made to study the same using a stochastic framework. 
In this connection, it is important to mention the theoretical modeling of bacterial 
two component system where stochastic kinetics has been used to study different 
phenotypic response (graded and all-or-none) \cite{Kierzek2010,Hoyle2012}.
In the present work, however, we have developed a mathematical 
formalism  to study signal transduction processes in generic bacterial TCS. While 
developing the model, we have taken into account only the post-translational 
modification in terms of phosphorylation-dephosphorylation kinetics as the timescale 
of the phosphotransfer kinetics is faster than the synthesis and/or degradation 
timescale of the system components \cite{Alves2003}. As mentioned earlier, the main 
role of TCS is to transmit the information of changes in the extra-cellular environment 
reliably within the cell. In the proposed stochastic study, we compare information 
processing in TCS with the monofunctional and the bifunctional property of sensor 
kinase. Combining both theoretical and numerical approaches, we show that for a 
fixed level of fluctuations due to stimulus, bifunctional TCS carries out a more reliable 
signal processing compared to monofunctional TCS.

The rest of the paper is organized as follows. In the next section, we develop the mathematical 
model to study signal transduction mechanism in monofunctional and in bifunctional system. 
Results of the model are discussed in Sec.~III and the paper is concluded in Sec.~IV.


\begin{figure}[!t]
\includegraphics[width=1.0\linewidth,angle=0]{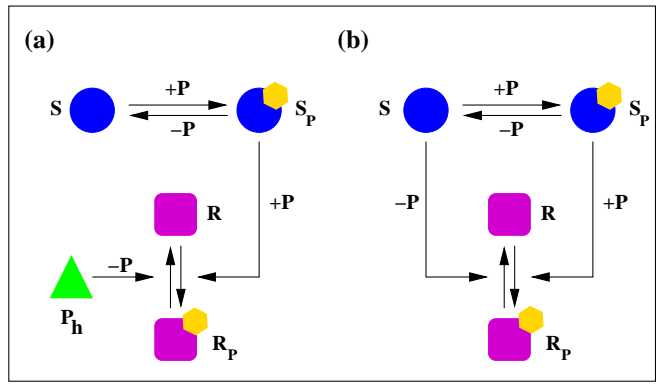}
\caption{(color online) Schematic diagram of phosphotransfer motif in (a) monofunctional 
and (b) bifunctional two component system. $S$ and $S_p$ stand for unphosphorylated 
and phosphorylated form of sensor kinase, respectively. Similarly, $R$ and $R_p$ stand 
for unphosphorylated and phosphorylated form of response regulator, respectively. $P_h$ 
stands for phosphatase. $\pm P$ stand for addition/removal of the phosphate group (shown 
by orange hexagon). Note that in monofunctional system, sensor kinase acts as a source of 
phosphate group whereas, in bifunctional system, it acts both as source and sink for the 
phosphate group.
}
\label{fig1}
\end{figure}


\section{The model}

Following phosphotransfer kinetics depicted in Fig.~(\ref{fig1}) for both monofunctional
and bifunctional systems, we have developed a mathematical model for noisy signal 
transduction in the present communication. In presence of an external inducer 
$I$, sensor kinase $S$ gets phosphorylated at the conserved histidine residue to
form $S_p$. Phosphorylated sensor then transfers the phosphate group to its cognate 
response regulator $R$, forming $R_p$. It is important to note that the above mentioned 
kinetics is common for both monofunctional and bifunctional systems. When it comes to 
removal of the phosphate group (dephosphorylation) from the response regulator, the two 
systems (monofunctional and bifunctional) behave differently. In the monofunctional system, 
the phosphate group from $R_p$  is removed by a phosphatase $P_h$ whereas, in the 
bifunctional one, the phosphate group is removed by the unphosphorylated
sensor $S$ itself. Thus, in monofunctional system, the sensor acts as a source of phosphate 
group and, in bifunctional system, the same acts as a sink in addition to being a source 
of the phosphate group. Considering the aforesaid interactions, the minimal kinetic 
steps for the phosphotransfer motif can be written as 
\begin{subequations}
\begin{eqnarray}
\label{eq1a}
\overset{k_{sI}}{\longrightarrow} I  & \overset{k_{dI}}{\longrightarrow} & \varnothing , \\
\label{eq1b}
S & \overset{k_p I}{\underset{k_{dp}}{\rightleftharpoons}} & S_p , \\
\label{eq1c}
S_p + R & \overset{k_{k}}{\longrightarrow} & S + R_p , \\
\label{eq1d} 
P_h + R_p & \overset{k_{pm}}{\longrightarrow} & P_h + R , \\
\label{eq1e}
\overset{k_{sP_h}}{\longrightarrow} P_h  & \overset{k_{dP_h}}{\longrightarrow} & \varnothing , \\
\label{eq1f} 
S + R_p & \overset{k_{pb}}{\longrightarrow} & S + R .
\end{eqnarray}
\end{subequations}

\noindent In the above kinetic steps, Eq.~(\ref{eq1a}) refers to the synthesis and 
degradation of the external inducer $I$. Eq.~(\ref{eq1b}) takes care of 
autophosphorylation and dephosphorylation of the sensor kinase. 
While modeling the autophosphorylation reaction, we have considered the signal $I$ as a 
catalyst which helps to convert the sensor $S$ to its phosphorylated form ($S_p$).
However, theoretical formalism developed earlier considered a more general framework
for stochastic signaling through enzymatic futile cycles \cite{Samoilov2005}.
Eq.~(\ref{eq1c}) considers the kinase reaction and Eq.~(\ref{eq1d}) considers the 
phosphatase activity of $P_h$ towards $R_p$. It is important to mention that while
writing the kinase and the phosphatase kinetics, we have considered second order
bi-molecular reaction scheme, although these reaction kinetics are generally
written using Michaelis-Menten type kinetics in the existing literature
\cite{Alves2003,Bandyopadhyay2012,Igoshin2008,Miyashiro2008,Tiwari2010}. 
One of the advantages
of using Michaelis-Menten kinetics is that it generates ultrasensitive switch in a
system \cite{Goldbeter1981,Goldbeter1982,TanaseNicola2006,Kierzek2010,Tiwari2010}
provided the network architecture generates substantial nonlinearity.
However, the reason behind using the second order bi-molecular reaction scheme 
in the present work is that it makes our analytical calculation tractable as shown in 
Sec.~IIA and Sec.~IIB.
Since the auxiliary protein $P_h$ behaves as an alternative source of phosphatase
in the monofunctional system, it is worthwhile to consider its kinetics (synthesis and
degradation) in the model. To this end, production and degradation of the 
phosphatase $P_h$ has been taken care of by Eq.~(\ref{eq1e}). 
Finally, 
Eq.~(\ref{eq1f}) is due to the phosphatase activity of $S$ towards $R_p$. Note that 
Eqs.~(\ref{eq1a}-\ref{eq1c}) are common for both monofunctional and bifunctional 
systems. Eqs.~(\ref{eq1d}-\ref{eq1e}) are exclusive for the monofunctional system and 
Eq.~(\ref{eq1f}) is solely for the bifunctional system. While writing the kinetic steps, we 
have mostly considered the post-translational modification for sensor and response
regulator. As mentioned earlier, we do not consider synthesis and degradation of the 
system components ($S$ and $R$) in the proposed model which keeps the total 
amount of sensor and response regulator constant, i.e., $S+S_p=S_T$ and $R+R_p=R_T$.

To understand the role of fluctuations prevalent due to the external inducer $I$ and the 
intrinsic cellular noise affecting the phosphotransfer mechanism, we adopt Langevin
approach to define different physical quantities. Langevin approach within the purview
of linear noise approximation is a valid approach provided fluctuations in the input 
signal are very small so that one can linearize the resultant noise in the Langevin 
equation \cite{Paulsson2004,Simpson2004,Shibata2005,TanaseNicola2006}. Such
linearization also remains valid when the coarse grained (steady state) time scale is 
longer than the birth-death rate of system components. In addition, a large copy number 
of system components makes the approximation valid. Since in TCS, copy numbers of 
$S$ and $R$ are large, one can adopt Langevin formalism to understand the stochastic
signal transduction mechanism. Thus, the Langevin equation associated with the inducer 
kinetics is given by,
\begin{equation}
\label{eq2}
\frac{dI}{dt} = k_{sI} - k_{dI} I + \xi_I ,
\end{equation}

\noindent where
\begin{equation}
\label{eq3}
\langle \xi_I (t) \rangle = 0 , \;
\langle \xi_I (t) \xi_I (t+\tau) \rangle = 2 k_{dI} \langle I \rangle \delta (\tau) ,
\end{equation}

\noindent with $\langle I \rangle$ being the mean inducer level at steady state. It is 
important to note that Eqs.~(\ref{eq2}-\ref{eq3}) are common for both monofunctional 
and bifunctional systems.

\subsection{Monofunctional system}

Considering the kinetic steps given by Eqs.~(\ref{eq1b}-\ref{eq1e}) and fluctuations 
associated with them, the Langevin equations for $S_p$, $R_p$ and $P_h$ for the 
monofunctional system can be written as
\begin{subequations}
\begin{eqnarray}
\label{eq4a}
\frac{dS_p}{dt} & = & k_p (S_T - S_p)  I - k_{dp} S_p \nonumber \\
&& - k_k S_p (R_T - R_p) + \xi_{S_p}, \\
\label{eq4b}
\frac{dR_p}{dt} & = & k_k S_p (R_T - R_p) - k_{pm} P_h R_p + \xi_{R_p}, \\
\label{eq4c}
\frac{dP_h}{dt} & = & k_{sP_h} - k_{dP_h} P_h + \xi_{P_h} .
\end{eqnarray}
\end{subequations}

\noindent The additive noise terms $\xi_{S_p}$, $\xi_{R_p}$ 
and $\xi_{P_h}$ take care of fluctuations in the copy number of $S_p$, $R_p$ and $P_h$, 
respectively. Using the concept of linear noise approximation, the statistical properties of
three fluctuating terms can be written as \cite{Kampen2005,Elf2003,Mehta2008}
\begin{subequations}
\begin{eqnarray}
\label{eq5a}
\langle \xi_{S_p} (t) \xi_{S_p} (t+\tau) \rangle_m & = & 2 k_p 
[ S_T - \langle S_p \rangle_m ] \langle I \rangle \delta (\tau) , \\
\label{eq5b}
\langle \xi_{R_p} (t) \xi_{R_p} (t+\tau) \rangle_m & = & 2 k_k  \langle S_p \rangle_m 
[ R_T - \langle R_p \rangle_ m ] \nonumber \\
&& \times \delta (\tau) , \\
\label{eq5c}
\langle \xi_{P_h} (t) \xi_{P_h} (t+\tau) \rangle_m & = & 2 k_{dP_h}  \langle P_h \rangle
\delta (\tau) ,
\end{eqnarray}

\noindent with $\langle \xi_{S_p} (t) \rangle_m = \langle \xi_{R_p} (t) \rangle_m = 
\langle \xi_{P_h} (t) \rangle_m = 0$. In the above equations, $\langle S_p \rangle_m$ and 
$\langle R_p \rangle_m$ stand for the mean values of $S_p$ and $R_p$ at the steady
state, respectively. Here, $\langle \cdots \rangle_m$ has been used to designate 
ensemble average for a monofunctional system. Furthermore, we consider that 
the noise terms 
$\xi_{S_p}$ and $\xi_{R_p}$ are correlated \cite{TanaseNicola2006,Swain2004}
\begin{equation}
\label{eq5d}
\langle \xi_{S_p} (t) \xi_{R_p} (t+\tau) \rangle_m  = - k_k \langle S_p \rangle_m 
[ R_T - \langle R_p \rangle_m ] \delta (\tau) .
\end{equation}
\end{subequations}

\noindent Linearizing Eq.~(\ref{eq2}) and Eqs.~(\ref{eq4a}-\ref{eq4c}) around the 
mean value at steady state, i.e.,  $I = \langle I \rangle + \delta I$, 
$S_p = \langle S_p \rangle_m + \delta S_p$,
$R_p = \langle R_p \rangle_m + \delta R_p$ and $P_h = \langle P_h \rangle + \delta P_h$, we have
\begin{eqnarray}
\label{eq6}
\frac{d}{dt}
\left (
\begin{array}{c}
\delta I \\
\delta S_p \\
\delta R_p \\
\delta P_h
\end{array}
\right )
& = &
\left (
\begin{array}{cccc}
- J_{I I}       &   J_{I S_p}       & J_{I R_p}           & J_{I P_h} \\
  J_{S_p I} & - J_{S_p S_p} & J_{S_p R_p}    & J_{S_p P_h} \\
  J_{R_p I} &   J_{R_p S_p} & - J_{R_p R_p} & - J_{R_p P_h} \\
  J_{P_h I}    &   J_{P_h S_p}    & J_{P_h R_p}      & - J_{P_h P_h}
\end{array}
\right
) 
\nonumber \\
&& \times
\left (
\begin{array}{c}
\delta I \\
\delta S_p \\
\delta R_p \\
\delta P_h
\end{array}
\right )
+
\left (
\begin{array}{c}
\xi_{I} \\
\xi_{S_p} \\
\xi_{R_p} \\
\xi_{P_h}
\end{array}
\right ) ,
\end{eqnarray}

\noindent with
\begin{eqnarray*}
&& J_{II} = k_{dI}, J_{I S_p} = J_{I R_p} = J_{I P_h} = 0 , \\
&& J_{Sp I} = k_p [ S_T - \langle S_p \rangle_m ] , \\
&& J_{S_p S_p}  = k_p \langle I \rangle + k_{dp} + k_k [ R_T - \langle R_p \rangle_m ] , \\
&& J_{S_p R_p} = k_k \langle S_p \rangle_m, J_{S_p P_h} = J_{R_p I} = 0 , \\
&& J_{R_p S_p} = k_k [ R_T - \langle R_p \rangle_m ] , \\
&& J_{R_p R_p} = k_k \langle S_p \rangle_m + k_{pm} \langle P_h \rangle, 
J_{R_p P_h} = k_{pm} \langle R_p \rangle_m , \\
&& J_{P_h I} = J_{P_h S_p} = J_{P_h R_p} = 0 .
\end{eqnarray*}

\noindent Fourier transformation 
($\delta \tilde{X} (\omega) = \int_{-\infty}^{+\infty}\delta X (t)\exp (-i\omega t)$ $dt$)
of Eq.~(\ref{eq6}) gives
\begin{eqnarray}
\label{eq6n}
\left (
\begin{array}{c}
i \omega \delta \tilde{I} \\
i \omega \delta \tilde{S}_p \\
i \omega \delta \tilde{R}_p \\
i \omega \delta \tilde{P}_h
\end{array}
\right )
& = &
\left (
\begin{array}{cccc}
- J_{I I}       &   J_{I S_p}       & J_{I R_p}           & J_{I P_h} \\
  J_{S_p I} & - J_{S_p S_p} & J_{S_p R_p}    & J_{S_p P_h} \\
  J_{R_p I} &   J_{R_p S_p} & - J_{R_p R_p} & - J_{R_p P_h} \\
  J_{P_h I}    &   J_{P_h S_p}    & J_{P_h R_p}      & - J_{P_h P_h}
\end{array}
\right
) 
\nonumber \\
&& \times
\left (
\begin{array}{c}
\delta \tilde{I} \\
\delta \tilde{S}_p \\
\delta \tilde{R}_p \\
\delta \tilde{P}_h
\end{array}
\right )
+
\left (
\begin{array}{c}
\tilde{\xi}_{I} \\
\tilde{\xi}_{S_p} \\
\tilde{\xi}_{R_p} \\
\tilde{\xi}_{P_h}
\end{array}
\right ) .
\end{eqnarray}

\noindent Solving Eq.~(\ref{eq6n}) yields
\begin{eqnarray*}
(i\omega + J_{S_p S_p}) \delta \tilde{S}_p
& = & J_{S_p I} \delta \tilde{I} + J_{S_p R_p} \delta \tilde{R}_p + \tilde{\xi}_{S_p} , \\
(i\omega + J_{R_p R_p}) \delta \tilde{R}_p
& = & J_{R_p S_p} \delta \tilde{S}_p - J_{R_p P_h} \delta \tilde{P}_h + \tilde{\xi}_{R_p} , \\
(i\omega + J_{I I}) \delta \tilde{I} & = & \tilde{\xi}_I , 
(i\omega + J_{P_h P_h}) \delta \tilde{P}_h = \tilde{\xi}_{P_h} , 
\end{eqnarray*}

\noindent which finally leads to the desired expression of $\delta \tilde{R}_p$ for 
the monofunctional system,
\begin{eqnarray}
\label{eq7}
\delta \tilde{R}_p & = & 
\frac{
J_{R_p S_p} J_{S_p I}  \delta \tilde{I} 
}{
(i\omega+J_{R_p R_p}) (i\omega+J_{S_p S_p}) - J_{R_p S_p} J_{S_p R_p}
}
\nonumber \\
&& -
\frac{
[ (i\omega+J_{S_p S_p}) J_{R_p P_h} ] \delta \tilde{P}_h 
}{
(i\omega+J_{R_p R_p}) (i\omega+J_{S_p S_p}) - J_{R_p S_p} J_{S_p R_p}
}
\nonumber \\
&& +
\frac{
( i\omega+J_{S_p S_p} )  \tilde{\xi}_{R_p}
}{
(i\omega+J_{R_p R_p}) (i\omega+J_{S_p S_p}) - J_{R_p S_p} J_{S_p R_p}
}
\nonumber \\
&& +
\frac{
J_{R_p S_p}   \tilde{\xi}_{S_p}
}{
(i\omega+J_{R_p R_p}) (i\omega+J_{S_p S_p}) - J_{R_p S_p} J_{S_p R_p}
} ,
\end{eqnarray}

\noindent
where
\begin{equation}
\label{eq8}
\delta \tilde{I} = \frac{\tilde{\xi}_I}{i\omega+J_{II}},
\delta \tilde{P}_h = \frac{\tilde{\xi}_{P_h}}{i\omega+J_{P_hP_h}}.
\end{equation}

\noindent In Eq.~(\ref{eq7}), the first term arises due to external inducer I. The second 
term is due to fluctuations in the phosphatase activity of $P_h$ on $R_p$. The third and 
fourth terms arise due to fluctuations in $R_p$ and $S_p$, respectively.
Now, using the expression of  $\delta \tilde{R}_p$ given in Eq.~(\ref{eq7})
and employing the properties of linear noise approximation given in 
Eqs.~(\ref{eq5a}-\ref{eq5d}), we define the variance associated with $R_p$ 
for the monofunctional system,
\begin{eqnarray}
\label{varm}
\sigma^2_{R_p} &=& \frac{1}{2\pi} \int d\omega 
\left \langle \left | \delta \tilde{R}_p (\omega) \right |^2 \right \rangle_m , \nonumber \\
&=&\frac{
J^2_{R_p S_p} J^2_{S_p I} \langle I \rangle (\alpha_m+\beta_m+J_{II})
}{
\alpha_m \beta_m (\alpha_m+\beta_m) (\alpha_m+J_{II}) (\beta_m+J_{II})
} \nonumber \\
&& + \frac{
J^2_{R_p P_h} \langle P_h \rangle
}{
\alpha_m \beta_m
} \nonumber \\
&& \times
\frac{
(\beta_m+J_{P_hP_h})J^2_{S_pS_p} + \alpha_m (\beta_m J_{P_hP_h}+J^2_{S_pS_p} )
}{
(\alpha_m+\beta_m) (\alpha_m+J_{P_hP_h}) (\beta_m+J_{P_hP_h})
} \nonumber \\
&& + \frac{
\gamma_m + 
k_k \langle S_p \rangle_m [R_T-\langle R_p \rangle_m] (J^2_{S_pS_p}+\alpha_m \beta_m)
}{
\alpha_m \beta_m (\alpha_m+\beta_m)
} ,
\end{eqnarray}

\noindent with
\begin{eqnarray*}
\alpha_m &=& 
\frac{1}{2} \left [ (J_{S_pS_p}+J_{R_pR_p}) \right. \\
&& \left. + \left \{ (J_{S_pS_p}-J_{R_pR_p})^2 + 4 J_{S_pR_p} J_{R_pS_p} \right \}^{1/2} \right ], \\
\beta_m &=& 
\frac{1}{2} \left [ (J_{S_pS_p}+J_{R_pR_p}) \right. \\
&& \left. - \left \{ (J_{S_pS_p}-J_{R_pR_p})^2 + 4 J_{S_pR_p} J_{R_pS_p} \right \}^{1/2} \right ], \\
\gamma_m &=&
J^2_{R_pS_p} k_p (S_T-\langle S_p \rangle_m) \langle I \rangle \\
&& - J_{R_pS_p} J_{S_pS_p} k_k \langle S_p \rangle_m (R_T-\langle R_p \rangle_m) .
\end{eqnarray*}

\subsection{Bifunctional system}

Considering the kinetic steps given by Eqs.~(\ref{eq1b}-\ref{eq1c},\ref{eq1f}) and 
fluctuations associated with them, the Langevin equations for $S_p$ and $R_p$ for 
the bifunctional system can be written as
\begin{subequations}
\begin{eqnarray}
\label{eq9a}
\frac{dS_p}{dt} & = & k_p (S_T - S_p)  I - k_{dp} S_p \nonumber \\
&& - k_k S_p (R_T - R_p) + \xi_{S_p}, \\
\label{eq9b}
\frac{dR_p}{dt} & = & k_k S_p (R_T - R_p) - k_{pb} R_p (S_T - S_p) \nonumber \\
&& + \xi_{R_p} .
\end{eqnarray}
\end{subequations}

\noindent The additive noise terms $\xi_{S_p}$ and  $\xi_{R_p}$ take care of fluctuations 
in the copy number of $S_p$ and $R_p$, respectively. The statistical properties of the 
two fluctuating terms are given by \cite{Kampen2005,Elf2003,Mehta2008}
\begin{subequations}
\begin{eqnarray}
\label{eq10a}
\langle \xi_{S_p} (t) \xi_{S_p} (t+\tau) \rangle_b & = & 2 k_p [ S_T - \langle S_p \rangle_b ] 
\langle I \rangle \delta (\tau) , \\
\label{eq10b}
\langle \xi_{R_p} (t) \xi_{R_p} (t+\tau) \rangle_b & = & 2 k_k  \langle S_p \rangle_b 
[ R_T - \langle R_p \rangle_ b ] \nonumber \\
&& \times \delta (\tau) ,
\end{eqnarray}

\noindent with $\langle \xi_{S_p} (t) \rangle_b = \langle \xi_{R_p} (t) \rangle_b = 0$. 
In the above equations, $\langle S_p \rangle_b$ and 
$\langle R_p \rangle_b$ stand for the mean values of $S_p$ and $R_p$ at the steady
state, respectively. Here, $\langle \cdots \rangle_b$ has been used to designate 
ensemble average for a bifunctional system. As in monofunctional system,
we consider the noise terms 
$\xi_{S_p}$ and $\xi_{R_p}$ to be correlated \cite{TanaseNicola2006,Swain2004}
\begin{equation}
\label{eq10c}
\langle \xi_{S_p} (t) \xi_{R_p} (t+\tau) \rangle_b  = - k_k \langle S_p \rangle_b 
[ R_T - \langle R_p \rangle_b ] \delta (\tau) .
\end{equation}
\end{subequations}

\noindent At this point it is important to note the difference between Eq.~(\ref{eq4b}) 
and Eq.~(\ref{eq9b}).
In Eq.~(\ref{eq4b}), the loss term ($k_{pm} P_h R_p$) appears due to phosphatase 
activity of $P_h$ on $R_p$, whereas in Eq.~(\ref{eq9b}), the loss term ($k_{pb} R_p (S_T - S_p)$)
appears due to phosphatase activity of $S$ on $R_p$. Although the noise term $\xi_{R_p}$ 
in both Eqs.~(\ref{eq4b},\ref{eq9b}) looks almost the same, it is the loss term in the aforesaid 
equations that makes the steady state behavior of $R_p$ different in monofunctional 
and bifunctional systems.
Now, linearizing as usual around the mean value at steady state we have
\begin{eqnarray}
\label{eq11}
\frac{d}{dt}
\left (
\begin{array}{c}
\delta I \\
\delta S_p \\
\delta R_p
\end{array}
\right )
& = &
\left (
\begin{array}{ccc}
- J_{I I}       &   J_{I S_p}       & J_{I R_p}  \\
  J_{S_p I} & - J_{S_p S_p} & J_{S_p R_p} \\
  J_{R_p I} &   J_{R_p S_p} & - J_{R_p R_p} \\
\end{array}
\right
) 
\nonumber \\
&& \times
\left (
\begin{array}{c}
\delta I \\
\delta S_p \\
\delta R_p
\end{array}
\right )
+
\left (
\begin{array}{c}
\xi_{I} \\
\xi_{S_p} \\
\xi_{R_p}
\end{array}
\right ) ,
\end{eqnarray}

\noindent with
\begin{eqnarray*}
&& J_{II} = k_{dI}, J_{I S_p} = J_{I R_p} = 0 , 
J_{Sp I} = k_p [ S_T - \langle S_p \rangle_b ], \\
&& J_{S_p S_p}  = k_p \langle I \rangle + k_{dp} + k_k [ R_T - \langle R_p \rangle_b ],
J_{S_p R_p} = k_k \langle S_p \rangle_b , \\
&& J_{R_p I} = 0, 
J_{R_p S_p} = k_k [ R_T - \langle R_p \rangle_b ] + k_{pb} \langle R_p \rangle_b , \\
&& J_{R_p R_p} = k_k \langle S_p \rangle_b + k_{pb} [ S_T - \langle S_P \rangle_b ] . 
\end{eqnarray*}

\noindent Fourier transforming Eq.~(\ref{eq11}) yields 
\begin{eqnarray}
\label{eq11n}
\left (
\begin{array}{c}
i\omega \delta \tilde{I} \\
i\omega \delta \tilde{S}_p \\
i\omega \delta \tilde{R}_p
\end{array}
\right )
& = &
\left (
\begin{array}{ccc}
- J_{I I}       &   J_{I S_p}       & J_{I R_p}  \\
  J_{S_p I} & - J_{S_p S_p} & J_{S_p R_p} \\
  J_{R_p I} &   J_{R_p S_p} & - J_{R_p R_p} \\
\end{array}
\right
) 
\nonumber \\
&& \times
\left (
\begin{array}{c}
\delta \tilde{I} \\
\delta \tilde{S}_p \\
\delta \tilde{R}_p
\end{array}
\right )
+
\left (
\begin{array}{c}
\tilde{\xi}_{I} \\
\tilde{\xi}_{S_p} \\
\tilde{\xi}_{R_p}
\end{array}
\right ) ,
\end{eqnarray}

\noindent
solution of which eventually leads to
\begin{eqnarray*}
(i\omega+J_{S_pS_p}) \delta \tilde{S}_p
&=& J_{S_pI} \delta \tilde{I} + J_{S_pR_p} \delta \tilde{R}_p +\tilde{\xi}_{S_p} \\
(i\omega + J_{R_p R_p}) \delta \tilde{R}_p
& = & J_{R_p S_p} \delta \tilde{S}_p + \tilde{\xi}_{R_p}.
\end{eqnarray*}

\noindent
Using the above two relations, we have the desired expression of $\delta \tilde{R}_p$
for the bifunctional system
\begin{eqnarray}
\label{eq12}
\delta \tilde{R}_p & = & 
\frac{
J_{R_p S_p} J_{S_p I}  \delta \tilde{I} 
}{
(i\omega+J_{R_p R_p}) (i\omega+J_{S_p S_p}) - J_{R_p S_p} J_{S_p R_p}
}
\nonumber \\
&& +
\frac{
( i\omega+J_{S_p S_p} ) \tilde{\xi}_{R_p}
}{
(i\omega+J_{R_p R_p}) (i\omega+J_{S_p S_p}) - J_{R_p S_p} J_{S_p R_p}
}
\nonumber \\
&& +
\frac{
J_{R_p S_p} \tilde{\xi}_{S_p}
}{
(i\omega+J_{R_p R_p}) (i\omega+J_{S_p S_p}) - J_{R_p S_p} J_{S_p R_p}
} ,
\end{eqnarray}

\noindent where the expression for $\delta \tilde{I}$ is given in Eq.~(\ref{eq8}). Now,
using the expression for $\delta \tilde{R}_p$ and properties of linear noise expression
given in Eqs.~(\ref{eq10a}-\ref{eq10c}), we write the variance associated with $R_p$ for
the bifunctional system
\begin{eqnarray}
\label{varb}
\sigma^2_{R_p} &=& \frac{1}{2\pi} \int d\omega 
\left \langle \left | \delta \tilde{R}_p (\omega) \right |^2 \right \rangle_b , \nonumber \\
&=&\frac{
J^2_{R_p S_p} J^2_{S_p I} \langle I \rangle (\alpha_b+\beta_b+J_{II})
}{
\alpha_b \beta_b (\alpha_b+\beta_b) (\alpha_b+J_{II}) (\beta_b+J_{II})
} \nonumber \\
&& + \frac{
\gamma_b + 
k_k \langle S_p \rangle_b [R_T-\langle R_p \rangle_b] (J^2_{S_pS_p}+\alpha_b \beta_b)
}{
\alpha_b \beta_b (\alpha_b+\beta_b)
} ,
\end{eqnarray}

\noindent with
\begin{eqnarray*}
\alpha_b &=& 
\frac{1}{2} \left [ (J_{S_pS_p}+J_{R_pR_p}) \right. \\
&& \left. + \left \{ (J_{S_pS_p}-J_{R_pR_p})^2 + 4 J_{S_pR_p} J_{R_pS_p} \right \}^{1/2} \right ], \\
\beta_b &=& 
\frac{1}{2} \left [ (J_{S_pS_p}+J_{R_pR_p}) \right. \\
&& \left. - \left \{ (J_{S_pS_p}-J_{R_pR_p})^2 + 4 J_{S_pR_p} J_{R_pS_p} \right \}^{1/2} \right ], \\
\gamma_b &=&
J^2_{R_pS_p} k_p (S_T-\langle S_p \rangle_b) \langle I \rangle \\
&& - J_{R_pS_p} J_{S_pS_p} k_k \langle S_p \rangle_b (R_T-\langle R_p \rangle_b) .
\end{eqnarray*}


\section{Results and discussion}

Since the main objective of the TCS signal transduction motif is to transduce the external
stimulus effectively and to generate the pool of phosphorylated response regulator $R_p$
that regulates several downstream genes, we now focus on quantifying different physical
quantities associated with $R_p$ for monofunctional and bifunctional systems.
While doing this, we make use of the expressions for $R_p$ given by Eq.~(\ref{varm}) and
Eq.~(\ref{varb}). Before proceeding further, it is important to mention the activity of kinase and 
phosphatase in monofunctional and bifunctional systems. In the monofunctional system, it has 
been observed that phosphatase has a higher affinity for the phosphorylated response regulator. 
On the other hand, in the bifunctional system, unphosphorylated sensor kinase has a lower 
affinity for the same \cite{Kenney2010}. Following these experimental information, kinase 
and phosphatase rate constants for both systems could be $k_k/k_{pm} < k_k/k_{pb}$. 
However, following earlier work on deterministic system, we consider $k_k/k_{pm} = 
k_k/k_{pb}$ as the particular parameter set as it has been 
shown to have a high degree of robustness \cite{Igoshin2008}. Furthermore, to check the 
validity of our proposed model, we perform stochastic simulation using Gillespie algorithm 
\cite{Gillespie1976,Gillespie1977} and find that the theoretical and numerical results are in 
good agreement with each other.

\begin{figure}[!t]
\begin{center}
\includegraphics[width=0.75\columnwidth,angle=-90,bb=48 62 590 792]{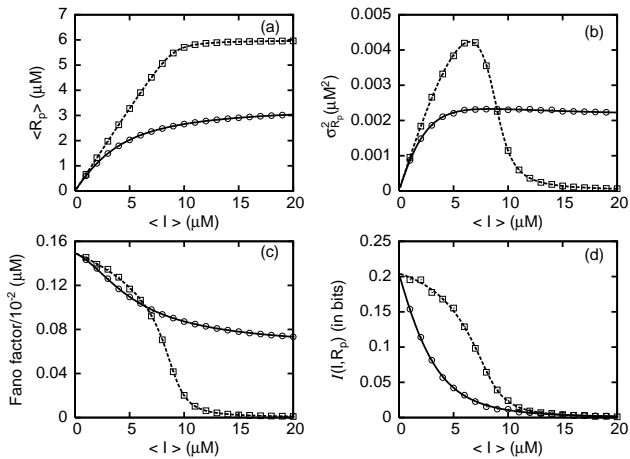}
\end{center}
\caption{Steady state
(a) $\langle R_p \rangle$ level,
(b) $\sigma^2_{R_p}$,
(c) Fano factor ($\sigma_{R_p}^2/\langle R_p \rangle$) and
(d) mutual information ${\cal I} (I,R_p)$
as a function of mean extra-cellular inducer level.
In all panels, solid (with open circles) and dashed (with open squares) lines are 
for monofunctional and bifunctional system, respectively.
The symbols are generated using stochastic simulation algorithm 
\cite{Gillespie1976,Gillespie1977} and the lines are due to theoretical calculation.
The parameters used are  \cite{Igoshin2008}
$k_p = 0.1$ $\mu$M$^{-1}$s$^{-1}$, $k_{dp} = 0.01$ s$^{-1}$,
$k_k = 0.2$ $\mu$M$^{-1}$s$^{-1}$, $k_{pb} = 0.15$ $\mu$M$^{-1}$s$^{-1}$,
$k_{pm} = 0.15$ $\mu$M$^{-1}$s$^{-1}$, $k_{sP_h} = 0.3$ $\mu$M s$^{-1}$,
$k_{dP_h} = 0.1$ s$^{-1}$, $S_T = 3.0$ $\mu$M and $R_T = 6.0$ $\mu$M.
}
\label{fig2}
\end{figure}

In Fig.~\ref{fig2}(a), we show the mean $R_p$ level at steady state,
$\langle R_p \rangle$, as a function of mean extra-cellular inducer level, 
$\langle I \rangle$. Initially, for low inducer level, both profiles grow linearly.
However, as inducer level increases, the profile of $\langle R_p \rangle$
for the monofunctional system (solid line) grows hyperbolically, whereas the 
same for the bifunctional system grows linearly (dashed line). It is important to 
mention that linear growth is a signature of linear input-output relation where 
the output level is dependent only on the input stimulus which increases the 
autophosphorylation rate in the model \cite{Shinar2007}. 
Linear input-output relation for the bifunctional system can be derived easily
from Eq.~(11), which provides the expression for mean $R_p$ level at steady 
state \cite{Miyashiro2008}
\begin{eqnarray*}
\langle R_p \rangle
& = & \frac{1}{2} \left ( R_T + \frac{k_{dp}}{k_k} + \frac{k_p \langle I \rangle}{k_{pb}} \right )
\nonumber \\
&& - \frac{1}{2} \sqrt{
\left ( R_T + \frac{k_{dp}}{k_k} + \frac{k_p \langle I \rangle}{k_{pb}} \right )^2
- \frac{4 k_p \langle I \rangle R_T}{k_{pb}}
} .
\end{eqnarray*}

\noindent
For $R_T > (k_{dp}/k_k) + (k_p \langle I \rangle/k_{pb})$,
we have $\langle R_p \rangle \approx (k_p/k_{pb}) \langle I \rangle$,
showing a linear relation between the input signal and the output.
On the other hand, using Eq.~(4), one can derive mean $R_p$ level at steady 
state for monofunctional system which takes into account both signals 
(input stimulus and phosphatase) as well as steady state value of mean $S_p$ 
\begin{eqnarray*}
\langle R_p \rangle =
\frac{
k_p \langle I \rangle S_T - ( k_p \langle I \rangle + k_{dp}) \langle S_p \rangle
}{
k_{pm} \langle P_h \rangle
} .
\end{eqnarray*}

\noindent
In this connection, it is important to mention that TetR-based negative autoregulation 
has been reported to linearize the dose-response (input-output) relation in 
\textit{S. cerevisiae} \cite{Nevozhay2009}.


Fig.~\ref{fig2}(a) further shows that for a fixed stimulus, amount of $R_p$ is always
higher for the bifunctional TCS. Thus, for a fixed stimulus, phosphotransfer mechanism
is more effective in producing a pool of $R_p$ for the bifunctional system compared 
to the monofunctional one and is in agreement with the result proposed earlier 
\cite{Alves2003}. Biologically, 
generation of a larger pool of $R_p$ is quite significant when it comes to the
phenomenon of gene regulation, as $R_p$ acts as a transcription factor for several
downstream genes. In the mechanism of gene regulation, a specific transcription factor
needs to attain a threshold value to make the genetic switch operative. Our result
suggests that for a target gene, a bifunctional system might work more effectively
than a monofunctional one by attaining the required pool of $R_p$ earlier.
Due to such a high activity, the bifunctional system will respond to a certain stimulus 
earlier than the monofunctional system by regulating downstream genes.

Fig.~\ref{fig2}(b) shows the profile of $\sigma^2_{R_p}$, variance of $R_p$. For
the monofunctional system, the variance grows steadily and then remains almost constant 
(solid line). However, the variance profile of the bifunctional system first grows to a maximum 
and then starts going down (dashed line). At a critical value of extra-cellular inducer 
level, almost all sensors and response regulators in the bifunctional system become 
phosphorylated, which, in turn, decrease fluctuations associated with $R_p$. Lowering 
of fluctuations in $R_p$ thereby reduces the variance. For the monofunctional case, in 
addition to the phosphorylation by the sensor kinase, an additional strong phosphatase 
activity is operational in the system which maintains sufficient fluctuations in the $R_p$ 
level, henceforth keeping the variance constant.

In the calculation of variance for the monofunctional system (see Eq.~(\ref{varm})), 
we have considered two extra sources of fluctuations, one due to the fluctuations 
in the kinetics of extra-cellular signal (Eq.~(\ref{eq1a})) and the other due to the 
fluctuations in the kinetics of phosphatase $P_h$ (Eq.~(\ref{eq1e})).
It is thus interesting to analyze whether fluctuations due to $P_h$ do
have any significant role in the variance of the monofunctional system. In the expression
of $\delta \tilde{R}_p$ for the monofunctional system (see Eq.~(\ref{eq7})), the second 
term appears due to the stochastic kinetics associated with the phosphatase $P_h$ 
(see Eq.~(\ref{eq4c})). However, for a constant level
of phosphatase, i.e., $\langle P_h \rangle$ one does not need to consider the 
stochastic kinetics given by Eq.~(\ref{eq4c}) that effectively removes fluctuations
associated with $P_h$ from both Eq.~(\ref{eq7}) and Eq.~(\ref{varm}). In addition, the
mean field contribution of $P_h$ appears in the second term on the right hand side
of Eq.~(\ref{eq4b}). Thus, for $\langle P_h \rangle$, the expression of variance for the 
monofunctional system becomes
\begin{eqnarray}
\label{varmc}
\sigma^2_{R_p} &=& \frac{
J^2_{R_p S_p} J^2_{S_p I} \langle I \rangle (\alpha_m+\beta_m+J_{II})
}{
\alpha_m \beta_m (\alpha_m+\beta_m) (\alpha_m+J_{II}) (\beta_m+J_{II})
} \nonumber \\
&& + \frac{
\gamma_m + 
k_k \langle S_p \rangle_m [R_T-\langle R_p \rangle_m] (J^2_{S_pS_p}+\alpha_m \beta_m)
}{
\alpha_m \beta_m (\alpha_m+\beta_m)
} .
\end{eqnarray}

\begin{figure}[!t]
\begin{center}
\includegraphics[width=0.75\columnwidth,angle=-90]{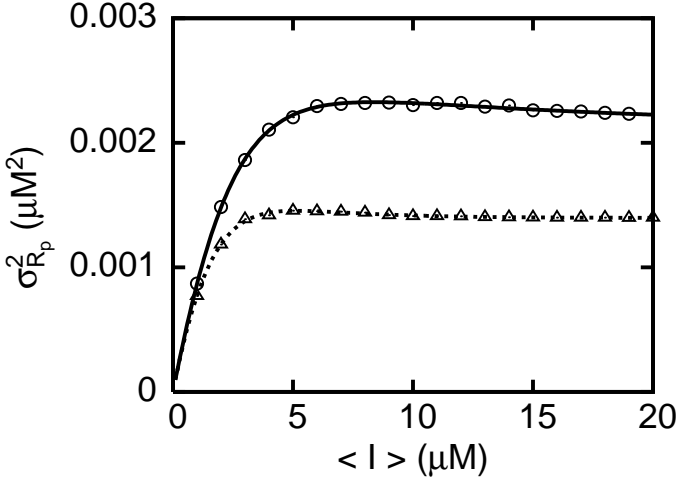}
\end{center}
\caption{Steady state variance $\sigma_{R_p}^2$ of monofunctional system.
The solid (with open circle) and dotted (with open triangle) lines are
due to fluctuating $P_h$ and constant $P_h$ ($\langle P_h \rangle$), respectively.
The symbols are generated using stochastic simulation algorithm 
\cite{Gillespie1976,Gillespie1977} and the lines are due to theoretical calculation.
For constant $P_h$, we have used $\langle P_h \rangle = 3$ $\mu$M.
The parameters used are same as in Fig.~\ref{fig2}
}
\label{fig3}
\end{figure}

\noindent
In Fig.~\ref{fig3}, we show variance associated with $R_p$ for the monofunctional 
system. The solid and dotted lines are due to presence and absence of fluctuations 
in $P_h$, respectively. It is clear from the profiles that for a constant level of 
phosphatase, $\sigma_{R_p}^2$ reduces appreciably compared to the fluctuating
$P_h$ level. This result suggests that stochastic kinetics of $P_h$ has a significant 
role in the fluctuations associated with $R_p$ in the monofunctional system.


\begin{figure}[!t]
\begin{center}
\includegraphics[width=0.75\columnwidth,angle=-90]{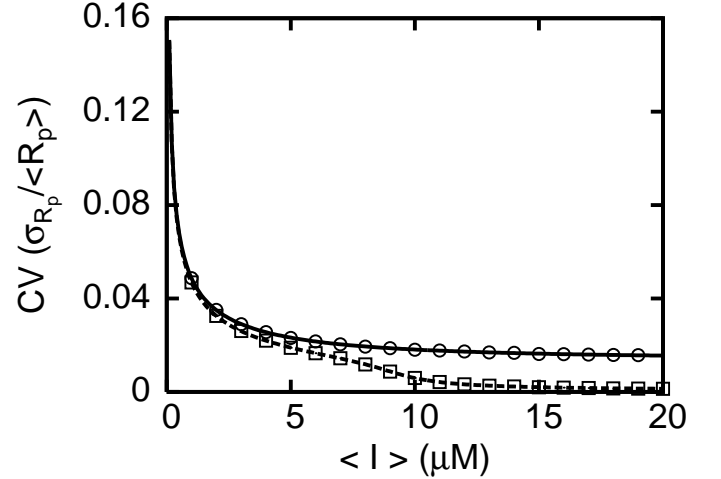}
\end{center}
\caption{Steady state CV ($\sigma_{R_p}/\langle R_p \rangle$) 
as a function of mean extra-cellular inducer level.
The solid (with open circles) and dashed (with open squares) lines are 
for monofunctional and bifunctional system, respectively.
The symbols are generated using stochastic simulation algorithm 
\cite{Gillespie1976,Gillespie1977} and the lines are due to theoretical calculation.
The parameters used are same as in Fig.~\ref{fig2}
}
\label{fig4}
\end{figure}

To quantify cellular fluctuations that affect phosphotransfer mechanism within the
TCS, we calculate Fano factor ($\sigma_{R_p}^2/\langle R_p \rangle$) 
\cite{Fano1947,Elf2003}
at steady state. In Fig.~\ref{fig2}(c), we have shown the profile of Fano factor for 
the monofunctional and the bifunctional systems (solid and dashed line, respectively) as a 
function of mean extra-cellular inducer level, where both profiles show decaying 
characteristics. Beyond a certain inducer level, Fano factor for the bifunctional system 
(dashed line) abruptly goes down to zero which can be attributed to the decaying nature 
of its variance shown in Fig.~\ref{fig2}(b). 
The pool of $R_p$ generated in the monofunctional system is not high enough to overcome 
fluctuations induced by the phosphatase $P_h$ and as a result, the fluctuations for this system 
maintain a low non zero value compared to the bifunctional system.
In addition to the Fano factor, we have also calculated the coefficient of variation (CV), i.e., 
$\sigma_{R_p}/\langle R_p \rangle$. In Fig.~\ref{fig4}, we show the steady 
state CV profile for both monofunctional and bifunctional systems.


As mentioned earlier, the specific job of a TCS is to sense change in the extra-cellular
environment and to transduce this information downstream reliably. To check how 
functionality of the sensor kinase affects fidelity (signal to noise ratio) of the signal 
processing mechanism, we calculate the quantity mutual information ${\cal I} (I,R_p)$ 
using the definition of Shannon 
\cite{Shannon1948,Borst1999}
\begin{eqnarray*}
{\cal I} (I,R_p) = \frac{1}{2} \log_2 \left ( 1+ \frac{\sigma^4_{IR_p}}{|C|} \right ) ;
C = \left (
\begin{array}{cc}
\sigma^2_{II} & \sigma^2_{IR_p} \\
\sigma^2_{IR_p} & \sigma^2_{R_pR_p}
\end{array}
\right ) ,
\end{eqnarray*}

\noindent where
\begin{eqnarray*}
\sigma^2_{II} = \langle I \rangle, 
\sigma^2_{IR_p} = \frac{
J_{R_pS_p} J_{S_pI} \langle I \rangle
}{
(\alpha_i + J_{II})(\beta_i + J_{II}),
}
\end{eqnarray*}

\noindent In the above relation, $i=m$ or $i=b$ depending on monofunctional and
bifunctional system, respectively. Expressions for 
$\sigma^2_{R_pR_p} (\equiv \sigma^2_{R_p})$  are given in Eqs.~(\ref{varm},\ref{varb}) 
for the two systems, respectively. Note that the quantity
$\sigma_{IR_p}^4/|C|$ stands for fidelity or signal to noise ratio 
\cite{Cheong2011,Bowsher2013}.

In Fig.~\ref{fig2}(d), we show mutual information profile for the monofunctional (solid line) 
and the bifunctional (dashed line) systems as a function of extra-cellular inducer level.
Information processing by both systems show a decaying profile as the extra-cellular
inducer level is increased. However, for a wide range of inducer level, information processing 
by the bifunctional system is higher than the monofunctional one. Beyond a certain
value of the inducer level, information profile of the bifunctional system goes down and
becomes equal to the profile of the monofunctional system. This result suggests that 
reliability of the bifunctional system in processing the information of the extra-cellular 
environment is higher than that of the monofunctional system.


\begin{figure}[!t]
\begin{center}
\includegraphics[width=0.75\columnwidth,angle=-90,bb=48 62 590 792]{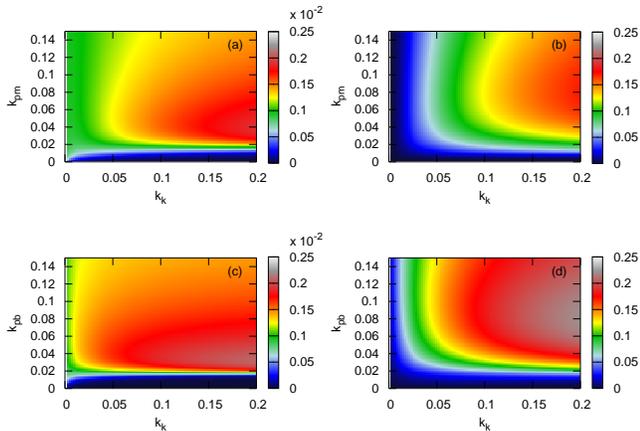}
\end{center}
\caption{(color online) Two dimensional map of Fano factor and mutual information, 
repectively, for monofunctional (a,b) and bifunctional system (c,d) as a function of kinase 
and phosphatase rate. The two dimensional map is a projection of Fano factor and mutual 
information on the kinase and phosphatase plane.
All panels are drawn using $\langle I \rangle$ = 1 $\mu$M. 
Values of other parameters are same as in Fig.~\ref{fig2}.
}
\label{fig5}
\end{figure}

To check the specific role of kinase and phosphatase rate on information 
processing within TCS further, we calculate Fano factor and ${\cal I} (I,R_p)$ 
as a function of kinase and phosphatase rate. 
In Figs.~\ref{fig5}(a)-\ref{fig5}(b), we show two dimensional map of Fano factor 
and mutual information, respectively, for monofunctional system as a function of 
$k_k$ and $k_{pm}$. Figs.~\ref{fig5}(c)-\ref{fig5}(d) show the same for bifunctional system 
as a function of $k_k$ and $k_{pb}$. Fig.~\ref{fig5}(a) shows that for high kinase 
and moderate phosphatase rate, the fluctuations level of phosphorylated response regulator 
becomes maximum otherwise it maintains a low value. This happens due to low 
copy number of proteins produced under high phosphatase regime. For the bifunctional 
system, the fluctuations level maintains a low value for a wide range of kinase and low 
phosphatase rate (see Fig.~\ref{fig5}(c)). Other than that, the fluctuations increase due to
increase in the phosphatase activity of the sensor protein. 
It is important to mention that maximum fluctuations level for the bifunctional system 
spans a wider region in the kinase-phosphatase plane compared to the
monofunctional system. Figs.~\ref{fig5}(b,d) 
show mutual information for monofunctional and bifunctional system, respectively. 
In both cases, signal processing capacity increases for high kinase and high 
phosphatase rate. In the regime of high kinase and high phosphatase activity, 
fluctuations in the $R_p$ copy number can reliably sense the fluctuations due to
extra-cellular stimulus which, in turn, effectively increases the signal processing 
capacity. In addition, due to structural advantage, information processing is better for 
the bifunctional system.


\begin{figure}[!t]
\begin{center}
\includegraphics[width=0.75\columnwidth,angle=-90,bb=48 62 590 792]{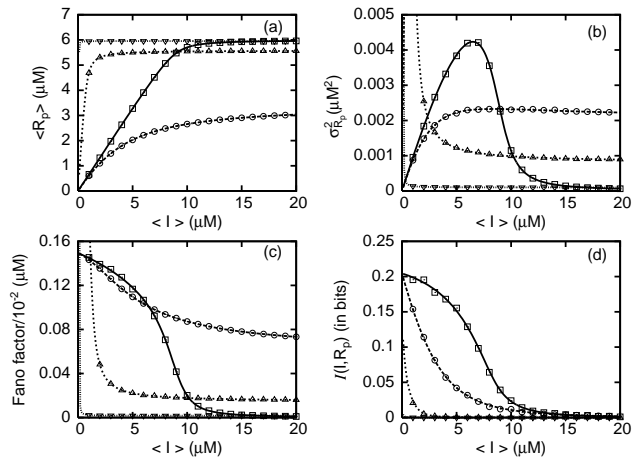}
\end{center}
\caption{Role of $P_h$ in (a) $\langle R_p \rangle$, (b) $\sigma_{R_p}^2$,
(c) Fano factor and (d) ${\cal I} (I,R_p)$ for monofunctional system. Open circle, up 
triangle and down triangle stand for $P_h = 3.0$, 0.3 and 0.03, respectively
(all in $\mu$M). 
Profile of bifunctional system (solid line with squares) has been shown in all 
four panels, for reference.
Values of the other parameters are same as in Fig.~(\ref{fig2}).
}
\label{fig6}
\end{figure}

Our analysis suggests that the bifunctional system can transduce external stimulus
more reliably than the monofunctional one. Effective signal transduction mechanism
of the bifunctional system can be attributed to its sensor domain which has synchronized
kinase and phosphatase activity. On the other hand, the monofunctional system lacks
such a synchronization due to the absence of phosphatase activity in the sensor domain.
At this point, one can ask the question, `Is it possible to increase the activity of a
monofunctional system by varying the contribution of one or more system components?' 
To answer this question we further looked into the signal transduction motif of 
the monofunctional system and found that due to the auxiliary protein $P_h$, the monofunctional 
TCS is unable to attain the activity of the bifunctional one. Using this phenomenological 
information, one may hypothesize that by reducing the effect of $P_h$, activity of 
the monofunctional system can be increased. To check this hypothesis, we have systematically 
reduced the concentration of $P_h$ by reducing the synthesis rate ($k_{sP_h}$) of $P_h$ from 
high to low value and calculated all physical quantities reported in Fig.~\ref{fig2} (see 
Fig.~\ref{fig6}). As the level of $P_h$ goes down, the phosphatase activity on the response 
regulator becomes more ineffective, and hence increases the pool of $R_p$ (see 
Fig.~\ref{fig6}(a)). For a low amount of $P_h$, $R_p$ level reaches the maximum value, 
as attained by the bifunctional system, quite early. This suggests that using the parameter set 
of the model and by lowering the amount of $P_h$, a monofunctional system can attain 
a large pool of $R_p$ even at a very low level of inducer. At a very low level of $P_h$, most 
of the response regulators get phosphorylated, henceforth reducing fluctuations in the 
$R_p$ level (see Figs.~\ref{fig6}(b),\ref{fig6}(c)). Interestingly, mutual information for 
the monofunctional system goes down with lowering of $P_h$ (see Fig.~\ref{fig6}(d)). By
reducing the level of $P_h$, fluctuations in $R_p$ level can be reduced, which, in turn, 
makes the system independent of fluctuations due to external stimulus. As a result, 
one observes suppression of mutual information.

Fluctuations due to inherent noisy biochemical reactions play an important role in gene
regulation by imposing phenotypic heterogeneity within genetically identical cells 
\cite{Kaern2005,Tiwari2010}. This happens due to fluctuations induced distribution of 
proteins in identical cells. In the present study, the TCS network output ($R_p$) shows 
maximal and minimal level of fluctuations for low and high stimulus, respectively. In 
addition to that, the bifunctional system maintains a lower noise profile compared to 
the monofunctional one. These results together suggest that the bifunctional system 
controlled gene regulation may have lesser variability (lower Fano factor) compared to the 
monofunctional system for intermediate to high stimulus level. To verify difference in 
variability in TCS controlled gene regulation, we consider a simple model of gene 
expression in the following section.

\subsection{TCS mediated gene regulation}

To understand the role of TCS on downstream genes, we consider a simple model of 
gene regulation mediated by either monofunctional or bifunctional TCS 
\begin{equation}
\label{gr1}
\overset{k_1 f(R_p)}{\longrightarrow} X  \overset{k_2}{\longrightarrow} \varnothing ,
\end{equation}

\noindent
where $X$ is the gene product whose synthesis is controlled by the transcription factor
$R_p$, output of TCS. The function $f(R_p)$ takes care of promoter switching
mechanism associated with the downstream gene and is given by $f(R_p)=R_p/(K+R_p)$,
with $K$ being the binding constant. While modeling promoter switching mechanism,
we have considered positive regulation by $R_p$. The stochastic differential equation 
associated with Eq.~(\ref{gr1}) is given by
\begin{equation}
\label{gr2}
\frac{dX}{dt} = k_1 f(R_p) - k_2 X + \xi_{X},
\end{equation}

\noindent
with $\langle \xi_X (t) \rangle = 0$ and 
$\langle \xi_X (t) \xi_X (t') \rangle = 2 k_2 \langle X \rangle \delta (t-t')$.
Fourier transformation of the linearized version of Eq.~(\ref{gr2}) yields
\begin{equation}
\label{gr3}
\delta \tilde{X} (\omega) = k_1 \left [ \frac{\delta f(R_p)}{\delta R_p} \right ]_{ss}
\frac{\delta \tilde{R}_p(\omega)}{(i\omega+k_2)} +
\frac{\tilde{\xi}_X (\omega)}{(i\omega+k_2)},
\end{equation}

\noindent
where $\left [\delta f(R_p)/\delta R_p \right ]_{ss} = K/(K+\langle R_p \rangle)^2$,
evaluated at steady state ($ss$).
Now, using expression of $\delta \tilde{X}$, we derive variance associated
with $X$,
\begin{equation}
\label{gr4}
\sigma^2_X = k_1^2 \left [ \frac{\delta f(R_p)}{\delta R_p} \right ]_{ss}^2
\frac{\sigma_{R_p}^2}{k_2} + \langle X \rangle .
\end{equation}

\noindent
Note that in Eq.~(\ref{gr4}), fluctuations due to the output of TCS is embedded
in the expression of $\sigma_{R_p}^2$, which is different for the monofunctional
and the bifunctional system (see Eq.~(\ref{varm}) and Eq.~(\ref{varb}), respectively).


\begin{figure}[!t]
\begin{center}
\includegraphics[width=0.75\columnwidth,angle=-90,bb=48 62 590 792]{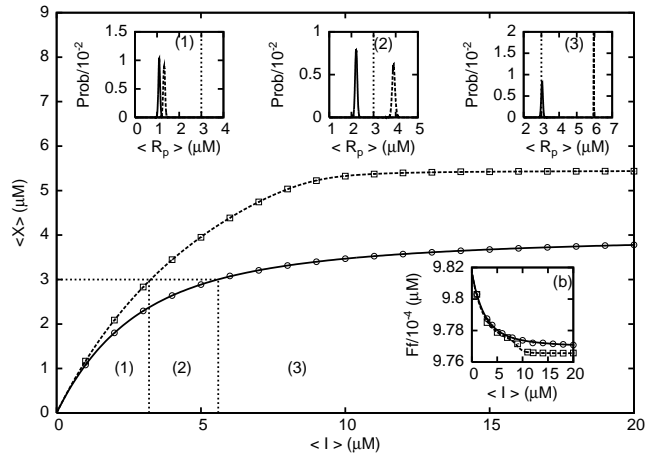}
\end{center}
\caption{Steady state $\langle X \rangle$ as a function of inducer 
$\langle I \rangle$. The solid (with open circle) and dashed (with 
open square) lines are $\langle X \rangle$ profile generated by monofunctional
and bifunctional system, respectively. Note that the lines and symbols are due
to theory and simulation \cite{Gillespie1976,Gillespie1977}, respectively.
The horizontal dotted line is for target $\langle X \rangle$ value
and the vertical dotted lines are corresponding inducer level for monofunctional
and bifunctional systems.
(1), (2) and (3) corresponds to three different regions of $\langle I \rangle$
for a fixed target $\langle X \rangle$.
Inset (b): Fano factor (Ff) for $X$ as a function of $\langle I \rangle$. Both lines and symbols
are same as in main figure.
Inset (1)-(3): Probability distribution of $R_p$ as a function of $\langle R_p \rangle$.
The solid and dotted lines represent probability distribution of
$R_p$ for monofunctional and bifunctional system, respectively.
The vertical dotted line denotes value of target $\langle X \rangle$ 
as shown in the main figure.
Parameters used are $K=5$ $\mu$M, $k_1 = 3 \times 10^{-3}$ $\mu$M$s^{-1}$,
$k_2 = 3 \times 10^{-4}$ $s^{-1}$ and values of the other parameters are same as 
in Fig.~\ref{fig2}.
}
\label{fig7}
\end{figure}

Main panel of Fig.~\ref{fig7} shows steady state $\langle X \rangle$ as a function of 
mean inducer level $\langle I \rangle$ controlled by monofunctional (solid line with open 
circle) and bifunctional system (dashed line with open square). It is evident from the profiles
of $\langle X \rangle$ that the bifunctional regulated system produces more downstream
protein as
$\langle I \rangle$ is increased. This happens due to availability of larger amount of
$\langle R_p \rangle$ produced by the bifunctional system compared to the monofunctional 
one, for a fixed amount of $\langle I \rangle$ (see Fig.~\ref{fig2}(a)). Fig.~\ref{fig7}(b)
depicts Fano factor ($\sigma^2_X/\langle X \rangle$) associated with downstream protein 
and shows that fluctuations in $X$ controlled by the bifunctional system is lower 
compared to the monofunctional system. Fluctuations in $X$ is controlled by fluctuations 
in $R_p$, as shown in Eq.~(\ref{gr4}). As fluctuations associated with $R_p$ due to
the bifunctional system is lower than the monofunctional system (see Fig.~\ref{fig2}(c)), its 
contribution in Fano factor associated with X is low.

In the previous discussion, we have shown that for a fixed level of inducer, a bifunctional 
system produces more $R_p$ compared to a monofunctional system (see 
Fig.~\ref{fig2}(a)). From this observation, we commented that a bifunctional system is 
more effective than a monofunctional one in regulating downstream genes for 
intermediate to high inducer level. To verify our remark, we numerically calculate 
probability distribution of $R_p$ (transcription factor for target gene) for monofunctional 
and bifunctional system (solid and dashed lines in inset (1)-(3) of Fig.~\ref{fig7})
for different values of $\langle I \rangle$ and check whether they are able to cross
the fixed $\langle X \rangle$ value (vertical dotted line in inset (1)-(3) of 
Fig.~\ref{fig7}). The distribution profiles give an idea of whether the pool of $R_p$ 
is able to generate a fixed level of gene product, in presence of inducer. For this, we 
set the value of downstream product $X$ at 3 $\mu$M (horizontal dotted line in the 
main panel of Fig.~\ref{fig7}) which intercepts both profiles of $\langle X \rangle$ at two 
different values of $\langle I \rangle$. To be explicit, for $\langle I \rangle = 3.2 \mu$M 
and $5.6 \mu$M, the horizontal dotted line intercepts the profile of bifunctional and 
monofunctional system, respectively. This, in turn, generates three different regions of 
$\langle I \rangle$, (1), (2) and (3) shown in the main panel of Fig.~\ref{fig7}. When 
value of $\langle I \rangle$ lies within region (1), the distribution profiles of $R_p$ for both 
systems are unable to cross the required value  of $\langle X \rangle$ (inset (1) of 
Fig.~\ref{fig7}). This scenario changes as we move to region (2). In this region, the 
distribution profile of the bifunctional system crosses target $\langle X \rangle$ value, but the
distribution profile of the monofunctional system is unable to cross the same (inset (2) 
of Fig.~\ref{fig7})). This happens due to low and high pool of $R_p$ generated by
monofunctional and bifunctional system, respectively. As we further move to region (3), 
we see that the distribution profiles of both the systems are able to cross the target $\langle X \rangle$ 
value as both produce enough $R_p$ to achieve the goal.


\section{Conclusion}

To conclude, we have developed a stochastic model for signal transduction mechanism 
in bacterial TCS. The proposed model takes into account the difference in the functionality
of the sensor kinase. This difference in functionality leads to a classification of the TCS,
viz., monofunctional and bifunctional. Considering only the phosphotransfer mechanism
within TCS triggered by external stimulus, we have derived Langevin equation associated 
with system components for both systems (monofunctional and bifunctional). Using the
expression of phosphorylated response regulators, we have calculated different physically 
realizable quantities, viz., variance, Fano factor (variance/mean), mutual information at steady 
state. Our analysis suggests that at low external stimulus, both systems reliably transduce
information due to changes made in the extra-cellular environment. Moreover, due to 
functional difference of the sensor kinase, it has been observed that fidelity of the bifunctional 
system is higher than that of the monofunctional system. Functionality of monofunctional 
system has been predicted to be increasable by reducing the amount of auxiliary protein 
($P_h$) which can be tested experimentally. We further extend our analysis by studying 
TCS mediated gene regulation which shows that the bifunctional system is more effective in 
producing target gene product for intermediate to high inducer level with lesser variability.


\begin{acknowledgments}
We thank Debi Banerjee for critical reading of the manuscript.
AKM and AB are thankful to UGC (UGC/776/JRF(Sc)) and CSIR (09/015(0375)/2009-EMR-I), 
respectively, for research fellowship. 
SKB acknowledges support from Bose Institute through Institutional Programme
VI - Development of Systems Biology.
\end{acknowledgments}


\end{document}